\documentclass[conference]{IEEEtran}
\usepackage{mathtools}
\usepackage{multirow}
\usepackage{graphicx}
\usepackage{subfigure}
\usepackage{wrapfig}
\usepackage{url}
\RequirePackage[bookmarks,
                urlcolor=black,
                citecolor=black,
                linkcolor=black,
                pagecolor=black,
                colorlinks,
                hyperfigures]{hyperref}
\usepackage{csquotes}
\usepackage{todonotes}
\usepackage[noend,linesnumbered]{algorithm2e}
\usepackage{placeins}
\usepackage[inline]{enumitem}
\usepackage{array}
\usepackage{listings}
\usepackage{pgfplots}
\usepackage{pgfplotstable}
\usepackage[gen]{eurosym}
\usepackage{amsfonts}
\usepackage{amsmath}

\newcommand*\circled[1]{\tikz[baseline=(char.base)]{\node[shape=circle,draw,inner sep=0pt] (char) {#1};}}

\newtheorem{definition}{Definition}

\begin{document}

\title{Temporal Conformance Checking at Runtime based on Time-infused Process Models}

\author{Florian Stertz, Juergen Mangler, Stefanie Rinderle-Ma}
\author{\IEEEauthorblockN{1\textsuperscript{st} Florian Stertz}
\IEEEauthorblockA{\textit{University of Vienna} \\
\textit{Faculty of Computer Science}\\
Vienna, Austria \\
florian.stertz@univie.ac.at}
\and
\IEEEauthorblockN{2\textsuperscript{nd} Juergen Mangler}
\IEEEauthorblockA{\textit{University of Vienna} \\
\textit{Faculty of Computer Science}\\
Vienna, Austria \\
juergen.mangler@univie.ac.at}
\and
\IEEEauthorblockN{3\textsuperscript{rd} Stefanie Rinderle-Ma}
\IEEEauthorblockA{\textit{University of Vienna} \\
\textit{Faculty of Computer Science}\\
Vienna, Austria \\
stefanie.rinderle-ma@univie.ac.at}
}

\maketitle

\begin{abstract} 	Conformance checking quantifies the deviations between a set of traces in a given process log and a set of possible traces defined by a process model.
Current approaches mostly focus on added or missing events. Lately, multi-perspective mining has provided means to check for conformance with time and resource constraints encoded as data elements. This paper presents an approach for quantifying temporal deviations in conformance checking based on infusing the input process model with a temporal profile. The temporal profile is calculated based on an associated process log considering task durations and the temporal distance between events. Moreover, a simple semantic annotation on tasks in the process model signifies their importance with respect to time. During runtime, deviations between an event stream and the process model with the temporal profile are quantified through a cost function for temporal deviations. The evaluation of the approach shows that the results for two real-world data sets from the financial and a manufacturing domain hold the promise to improve runtime process monitoring and control capabilities.

\end{abstract}

\section{Introduction}
\label{Sec:Intro}

The current pandemic has put a spotlight on the ability of companies and organizations to react quickly on a radically changed situation.
At the same time, business processes have (re-)gained tremendous interest due to their key role in driving digitalization.
The consequence is that companies and organizations have to be able to deal with change and evolution in their business processes. 

The intended behavior of a business process is described by a \textsl{process model}. Based on the model, during runtime, \textsl{process instances} are created and executed. 
Information on the execution of the process instances is stored in so called \textsl{process execution logs}, usually defined in the eXtensible Event Stream (XES) format \cite{7740858}. For each of the process instances, a \textsl{trace} is stored, reflecting the sequence of \textsl{events} that occurred for the process instance. An event contains information about the executed activities, and an arbitrary number of data elements, i.e., a timestamp or resource. The process execution log hence reflects the actual behavior of the process. 

\textsl{Conformance checking}, one of the three main areas of process mining \cite{DBLP:books/sp/Aalst16},
aims at quantifying the conformance between the described and actual behavior of a process
\cite{DBLP:books/sp/CarmonaDSW18}. For this, the process model and a process execution log are analyzed for deviations. If no deviations can be found (``perfect'' conformance), a fitness score of $1$ is assigned. For any mismatches (i.e., missing events, added events) the fitness score is reduced.

So far, existing approaches have mainly focused on control flow conformance. Multi perspective process mining denotes a research direction that emphasizes the importance of considering other process perspectives such as data, time, and resources for conformance checking, as well \cite{DBLP:journals/computing/MannhardtLRA16}.





This paper will pick up this line of argumentation and, in addition to control flow conformance, focus on \textsl{temporal deviations} in conformance checking. 
Temporal information in a process can be
defined in two different ways: (a) the time between start events of two
subsequent tasks (\textsl{temporal distance}) and (b) the time between start and end events of a single task (\textsl{task duration}) \cite{DBLP:journals/sosym/PosenatoLCR19}.
In general, logs for mining purposes are either generated by extracting data
from a specialized information system, or alternatively generated by a process
execution engine. Most available data sets only contain end events. Thus
temporal distance is the prevalent definition, while some process managements systems produce
logs with start and end events and thus allow for the more concise task duration point of view.

Temporal deviations might occur for the following reasons:

\begin{itemize}

  \item Reduced duration: A task was not executed properly because, e.g., a machine failed or
  information was missing. 

  \item Increased duration: Preconditions for the tasks have not been met, or
  hidden dependencies between tasks exist.

\end{itemize}




This paper aims at finding and quantifying such temporal deviations. By contrast to existing work \cite{DBLP:journals/computing/MannhardtLRA16}, the temporal information is not available in the form of temporal constraints, but is determined based on an input process execution log. We call the result a \textsl{temporal profile}. The input process model is infused with the temporal profile. 
The goal of this work is to determine temporal deviations of an (ongoing) process event stream with the temporal profile during runtime (online). A \textsl{process event stream} contains
events reflecting the execution of process instances. Instead of a protocol of finished process
instances, the events of all process instances currently being executed are
being put into a stream. An event stream contrary to a process execution log is
infinite.








Another contribution of this work is to propose a user-adjustable semantic quantification (in the form of
an annotation to the process model) in order to quantify the significance of temporal deviations for a certain task. This way, it can be expressed that exceeding the limit of a task
duration can be more severe for some events (or tasks) and even affect
succeeding events.


Two real-world data sets are used for the evaluation of the approach. The first one is the BPI Challenge 2012, since it meets the criteria of providing
the start and end time of at least some events of the process execution log \cite{bpic12}. The other data set is from the manufacturing domain and features the production
of industry parts where the whole production process is managed and enacted by
a process execution engine.

In Section \ref{Sec:fund}, fundamentals on
conformance checking are sketched. Section \ref{Sec:cont} provides the algorithm for infusing process model with a temporal profile. Moreover, Sec. \ref{Sec:cont} introduces the
 cost function for temporal deviations and the algorithm
incorporating it. The contribution is then evaluated and discussed in Sec.
\ref{Sec:eval}. Related work in this field is presented in Sec. \ref{Sec:rel}
Section \ref{Sec:concl} presents conclusions and future work.

\section{Fundamentals}
\label{Sec:fund}

\begin{figure}[htb!]
\centering
\includegraphics[width=0.6\columnwidth]{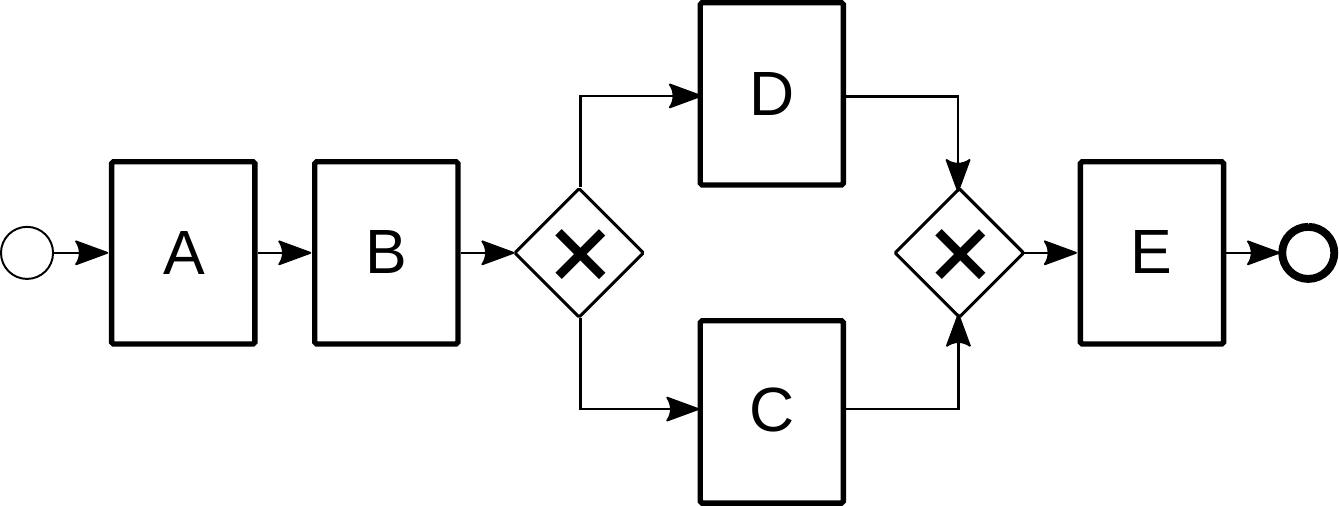}
\caption{Process model containing 5 tasks and 1 decision gateway}\label{fig:fundi_ex}
\end{figure}

This section provides a brief introduction into conformance checking 
\cite{DBLP:books/sp/Aalst16}. 
Figure \ref{fig:fundi_ex} shows an abstract example of
a process model. Task $A$ is followed by task $B$. Either task $C$ or $D$ can be executed following $B$.
After either of $C$ or $D$ finished, task $E$ can be executed.



Conformance checking is one of the three main areas of process mining. It determines
if the behavior of a process instance fits the description of a process model. Originally
token based replay has been used on a petri net representation of the process model to determine the conformance. Lately, aligning the event sequence of the process instance to the process model is used \cite{van2012replaying,adriansyah2014aligning}.

\begin{table}[htb!]
\centering
\begin{tabular}{l|lllll}
Model & A & B & D     & E & $\gg$ \\ \hline
$t_3$ & A & B & $\gg$ & E & D
\end{tabular}
\vspace{2mm}
\caption{Alignment with two moves due to wrong order.}
\label{tab:cc}
\end{table}

Based on the process model in Fig. \ref{fig:fundi_ex}, the following traces can be generated: 
$t_1$ = ($A$,$B$,$C$,$E$) and $t_2$ =
($A$,$B$,$D$,$E$). $t_1$ and $t_2$ match the process model perfectly.
Given a trace $t_3$ = ($A$,$B$,$E$,$D$), a deviation can be detected. Event $E$
and $D$ appear in the wrong order. To align $t_3$ to the process model,
so called asynchronous moves are added to the \textsl{alignment}, depicted with a $\gg$
in the alignment. The alignment necessitates two moves, as can be
seen in Tab. \ref{tab:cc}. This is not the only possible alignment for $t_3$,
as a move in the model and then in the log can be possible as well. To
calculate the best alignment for a trace a cost is added for every asynchronous
move. Perfectly matching traces yield a cost of $0$, while every asynchronous
move can increase this value. The cost value for a asynchronous move is defined
in a \textsl{cost function}. Usually every asynchronous move is assigned a cost value of
$1$, resulting in an alignment cost of $2$ for $t_3$. Conformance checking aims to
find the alignment with the minimum cost.

\section{Temporal Conformance Checking}
\label{Sec:cont}

This section 
infuses process models with temporal profiles as input for temporal conformance checking using a cost function to quantify temporal deviations. 

\subsection{Temporal Profile Generation}


The basic idea of temporal conformance checking is to infuse a process model
with a temporal profile and then to conduct conformance checking using a cost
function that quantifies temporal deviations between the temporal profile and
an event stream of interest. The input for temporal conformance checking hence
comprises a process model and (i) a process log in an offline setting or (b) an
event stream in a runtime (online) setting. For (i) the process log can be
split into a training set for calculating the temporal profile and a test set
for temporal conformance checking. The evaluation will show both, (i) offline
and (ii) online settings.

The temporal profile captures task duration and temporal distance between
events/tasks. We will explain both in the following and sketch some scenarios.
Algorithm \ref{alg:a1} calculates a temporal profile from a process model and a
process execution log. An example for a process model infused with a temporal
profile is provided in Sec. \ref{sub:example}. 

\begin{figure}[htb!]
\centering
\includegraphics[width=0.35\textwidth]{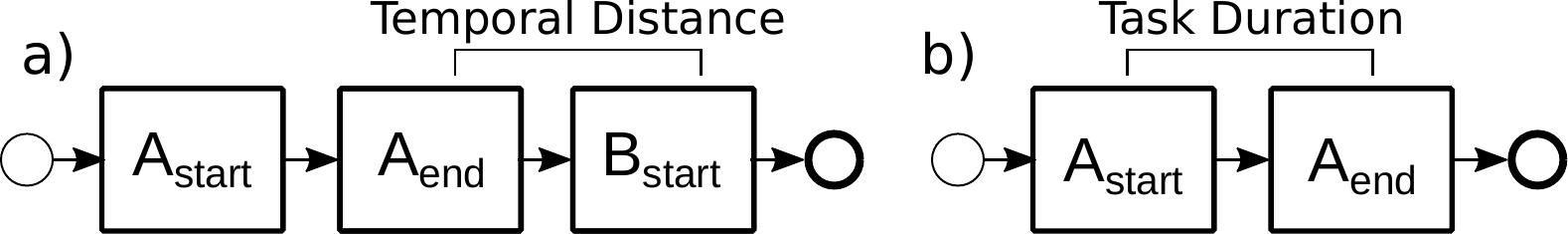}
\caption{Temporal Profile -- Example.}\label{fig:inter_ex}
\end{figure}

\textbf{Task Duration}: 
To compare the attached event data between process execution logs in the XES format,
standard extensions are available, for example, the name, the timestamp, and
the lifecycle of an event. The lifecycle data element of an event reflects its
current status. For temporal conformance checking, start and end lifecycle
event are needed to calculate the complete task duration by calculating
the difference between these two timestamps (cf. Fig. \ref{fig:inter_ex} (b)).


\textbf{Temporal Distance}: The temporal distance determines
the time after an event is completed and before a new event starts. For this, the
events have to contain again a data element describing its lifecycle, i.e.,
start and complete. We use the notation $|AB|$ for describing the temporal distance between event \texttt{A} and \texttt{B}.  While in a strict
sequence of events, like events \texttt{A} and \texttt{B} in Fig.
\ref{fig:inter_ex} (a), the computation of the temporal distance is
trivial, some interesting occurrences can be witnessed if events are executed
in parallel, i.e., event \texttt{C} and \texttt{D} in Fig. \ref{fig:running_ex}.

Event \texttt{B} has to be completed before either event \texttt{C} or
\texttt{D} can be started. This leads to 4 possible temporal distances,
namely $|BC|$, $|BD|$, $|CD|$, and, $|DC|$. Both, \texttt{C} and \texttt{D}
have to be finished before event \texttt{E} can be performed, thus the
distances $|DE|$ and $|CE|$ emerge as well.

Algorithm \ref{alg:a1} calculates the mean and standard deviation for all observed temporal distances and stores them.
Infrequent distances can be filtered out using a threshold value.

\begin{algorithm}[htb!]
\footnotesize
\SetAlgoCaptionSeparator{\footnotesize .}
\SetAlgorithmName{\footnotesize Alg.}{}

 \KwIn{\textbf{$M$}: A process model,\\
 \textbf{$L$}: A process execution log, containing traces of process instances,
 \textbf{$\kappa$}: threshold for filtering infrequent time distances
  }
  \KwResult{\textbf{$M$}: A process model containing information of time distances  }

  $td$ = dict()  // Hash Map of Execution time duration

  $inter\_td$ = dict()  // Hash Map of Inter Time distances

  \For{trace in $L$}{ \label{a1:alloop}  // iterate over every process instance in the log

    $current\_starts$ = dict() // Hash Map for execution time duration

    $last\_end$ = None

    \For{event in trace}{ //iterate over every event of trace

      // lc() == lifecycle of event, ts() == timestamp of event

      \If{event.lc() == 'start'}{

        $current\_starts$[event.name] = event.ts()

        \If{$last\_end$ != None}{

          \If{$inter\_td$[$last\_end$.name+event.name] == None}{

            $inter\_td$[$last\_end$.name+event.name] = list()

          }

          $inter\_td$[$last\_end$.name+event.name].append( $last\_end$.ts()-event.ts())

        }

      }

      \If{event.lc() == 'complete'}{

        \If{$td$[event.name] == None}{

          $td$[event.name] = list()

        }

        $td$[event.name].append(event.ts()-$current\_starts$[event.name].ts())

        $current\_starts$.remove(event.name)

        $last\_end$ = event

      }

    }

  }

  $stats$ = dict()   // Hash Map of all means and standard deviations of time distances

  \For{key in $td$}{ \label{a1line:td}

    $stats$[key] = (mean($td$[key]),stddev($td$[key]))

  }

  \For{key in $inter\_td$}{ \label{a1line:inter}

    \If{len($inter\_td$[key]) $\geq \kappa$}{

      $stats$[key] = (mean($inter\_td$[key]),stddev($inter\_td$[key]))

    }

  }

  $M$.add($stats$) // Infusing Process Model with Hash Map of time distances

  return $M$
\caption{\footnotesize Temporal Profile Generation}
\label{alg:a1}
\end{algorithm}

For this, Alg. \ref{alg:a1} starts with creating two hash tables \cite{cormen2009introduction} to save the time
distances, one for task duration and one for temporal distances.
Afterwards, each event of every process instance is parsed. If a start event is
detected, the temporal distance is calculated to the last detected end event
and the timestamp is saved for the task duration. On the other hand,
if an end event is detected, the task duration is calculated and the
event is saved as the currently last finished event for the temporal
distance. Lines \ref{a1line:td} and \ref{a1line:inter} determine the mean and
standard deviation and filter out infrequent temporal distances.
If no suitable data set is available, domain experts could produce the temporal profile.

\subsection{Temporal Conformance Checking}

After temporal profile is
calculated, temporal conformance checking can be performed, either online
at runtime on an event stream or offline on a process execution log.

The \texttt{z-score} \cite{crocker1986introduction} is used to determine the distance of new observations to
the gathered data sets from the preparation phase. Since the data for the data sets
has been collected in a complete test set, it can therefore be argued that the data set
is complete, which enables the use of the z-score. The z-score is defined as follows:

\[
  z = |\frac{x-\mu}{\sigma}|
\]

Thus, the deviation is calculated by subtracting the mean of all time distances
for an event from the new observation and divides it by standard deviation of
all time distances for one event. If the z-score exceeds a specified threshold,
a deviation is detected. Certain tasks allow a greater deviation while other
tasks demand to be very precise. The z-score determines how many standard
deviations the observation is distant to the mean. Thus a list of thresholds is
used, containing a threshold for every task in the process model to reflect the
needs for adjustable outlier detection.

To use this z-score for quantifying the cost of a temporal deviation, a temporal deviation
cost is introduced as follows:

\begin{figure*}
\begin{definition}[Cost Function for Temporal Deviations]\label{def:cost}
  \[
    temporal\_deviation\_cost\_function(x,M)=
  \begin{cases}
    0, & \text{if no $mean(M_x)$ available $||$ z-score(x) $\leq \kappa_{M_x}$ } \\
    \omega_{M_x} * \phi * \text{z-score}(x) & \text{otherwise} \\
  \end{cases}
  \]
  \\
  Let $x$ be a time distance, $M$ a process model containing information of time distances,
  $\kappa$ a threshold defined between 0 and $\infty$
  and $M_x$ the data set of related time distances, containing the mean and
  standard deviation and a related $\kappa$-threshold. If a mean is available and the z-score is smaller or
  equals $\kappa$, this function yields 0.
  Otherwise let $\phi$ any number between 0 and $\inf$ to adjust
  the impact of a temporal deviation in general and let $\omega_{M_x}$
  be a weight between 0 and $\infty$ for the related event to adjust the impact
  of specific events. The function yields the temporal deviation cost for observations
  with a z-score greater than the threshold $\kappa$.
\end{definition}
\end{figure*}

The end cost for an alignment calculated by temporal deviation conformance checking
is the sum of all costs of structural movements using standard conformance
checking, i.e., log and model moves, and of all costs of temporal deviations using
the temporal deviation cost function.

\begin{algorithm}[htb!]
\footnotesize
\SetAlgoCaptionSeparator{\footnotesize .}
\SetAlgorithmName{\footnotesize Alg.}{}

  \KwIn{\textbf{$M$}: Process Model with information on time distances\\
 \textbf{$ES$}: An event stream, sending events related to model $M$. \\
 \textbf{$TSIZE$}: Maximum Number of available process instances that can be stored. \\ 
 \textbf{$\kappa_{event}$} a list of thresholds the maximum allowed z-score per event \\
 \textbf{$\omega_{event}$} a list of weights for the results of temporal deviation cost function  per event \\
 \textbf{$\phi$}: general cost modifier for temporal deviations
  }
  \KwResult{\textbf{$C$}: Cost for last alignment of $ES$.}
  // temporal\_deviation\_cost\_function(x) == tc(x)

  $traces$ = dict() \label{a2:tradict}

  \For{$e$ in $ES$}{

    \If{$e$.trace not in $traces$}{ 
      
      \If{len($traces$) $\geq$ $TSIZE$}{ \label{a2:remove}

        $traces$.remove\_oldest()
      
      }
      
      $traces$[$e$.trace] = dict()  \label{a2:stuff}

      $traces$[$e$.trace]['$cost\_time$] = 0

      $traces$[$e$.trace]['$preceding\_event$] = None

      $traces$[$e$.trace]['$unfinished\_events$] = dict()

      $traces$[$e$.trace]['trace'] = list() \label{a2:stuffe}

    }

    $t$ = $traces$[$e$.trace]['trace']

    $t$.append($e$)

    $traces$[$e$.trace]['$cost\_structural$']= online\_conformance\_checking($t$,$e$) \label{a2:cc}

    \If{$e$.lc() == 'complete'}{

      \If{tc($e$,$M$) $> \kappa$}{

        $traces$[$e$.trace]['$cost\_time$'] += tc($e$,$\omega_{e}$, $\phi$) \label{a2:dur}

      }

      $traces$[$e$.trace]['$preceding\_event$'] = $e$

      $traces$[$e$.trace]['$unfinished\_event$'].remove($e$.name) \label{a2:remove}

    }

    \If{$e$.lc() == 'start'}{

      \If{$traces$[$e$.trace]['$preceding\_event']$ != None and tc($|$$traces$[$e$.trace]['$preceding\_event$']$e$$|$) $> \kappa_{e}$}{

        $traces$[$e$.trace]['$cost\_time$'] += tc($|$$traces$[$e$.trace]['$preceding\_event$']$e$$|$,$\omega_{e}$, $\phi$) \label{a2:inter}
      }

      $traces$[$e$.trace]['$unfinished\_events$'][$e$.name] = $e$.ts()

    }

    \For{event in $traces$[$e$.trace]['$unfinished\_events$']}{

      \If{Time.now - $M$[event].mean() $>$ 0 and tc(event(Time.now-$traces$[$e$.trace]['$unfinished\_events$'][event]),$\omega_{e}$, $\phi$) $> \kappa_{e}$}{

        $traces$[$e$.trace]['$cost\_time$'] += tc(event(Time.now-$traces$[$e$.trace]['$unfinished\_events$'][event]),$\omega_{event}$, $\phi$) \label{a2:unf}

      }

    }
  }
    $C$ = $traces$[$e$.trace]['$cost\_structural$'] + $traces$[$e$.trace]['$cost\_time$']

  return $C$
\caption{\footnotesize Finding Cost of Alignment}
\label{alg:a2}
\end{algorithm}

Algorithm \ref{alg:a2} shows a prototypical implementation for temporal
conformance checking at runtime.  Line \ref{a2:tradict} starts a hash table for
all currently saved process instances. The parameter $TSIZE$ defines how many
process instances can be stored at the same time. Since we are dealing with
possibly infinite process instances, only a fixed number can be stored.  If the
hash table is full, an older process instance is removed from the table. There
are many classifiers possible for determining the to be removed process
instance. This approach opted for the oldest process instance to be removed,
line \ref{a2:remove}. Lines \ref{a2:stuff} to \ref{a2:stuffe} create the
necessary variables for each instance.

In line \ref{a2:cc} the alignment cost for standard conformance checking is
calculated \cite{van2017online}. The hash map $unfinished\_events$ saves
all timestamps of starting events. Every time a starting event is detected, the
timestamp is added. If the related end event is detected, line \ref{a2:remove},
the timestamp is removed from the hash map. For each end event, the temporal
deviation cost is calculated based on the duration of the event, line
\ref{a2:dur}. For every detected start event, the temporal deviation cost for inter
event time distance is calculated, line \ref{a2:inter}.

If an event is still being executed and its task duration is already greater
than the average time for this task, the temporal deviation cost is calculated
using the time duration between the moment the last event is detected in the
event stream of this process instance and the starting event, line
\ref{a2:unf}. Instead of event for events in the event stream, the algorithm
can be applied periodically as well to update the temporal deviation cost for
unfinished events.  As can be seen in Alg. \ref{alg:a2},  the temporal
deviation cost is calculated in linear time, leaving calculating the structural
conformance checking score, Line \label{a2:cc}, as the only heavy computing
task. Splitting the computation of the structural and temporal deviation costs
can yield a performance increase and allows for a scalable application of
temporal conformance checking.

\subsection{Illustrating Example}
\label{sub:example}

\begin{figure}[htb!]
\centering
\includegraphics[width=0.85\columnwidth]{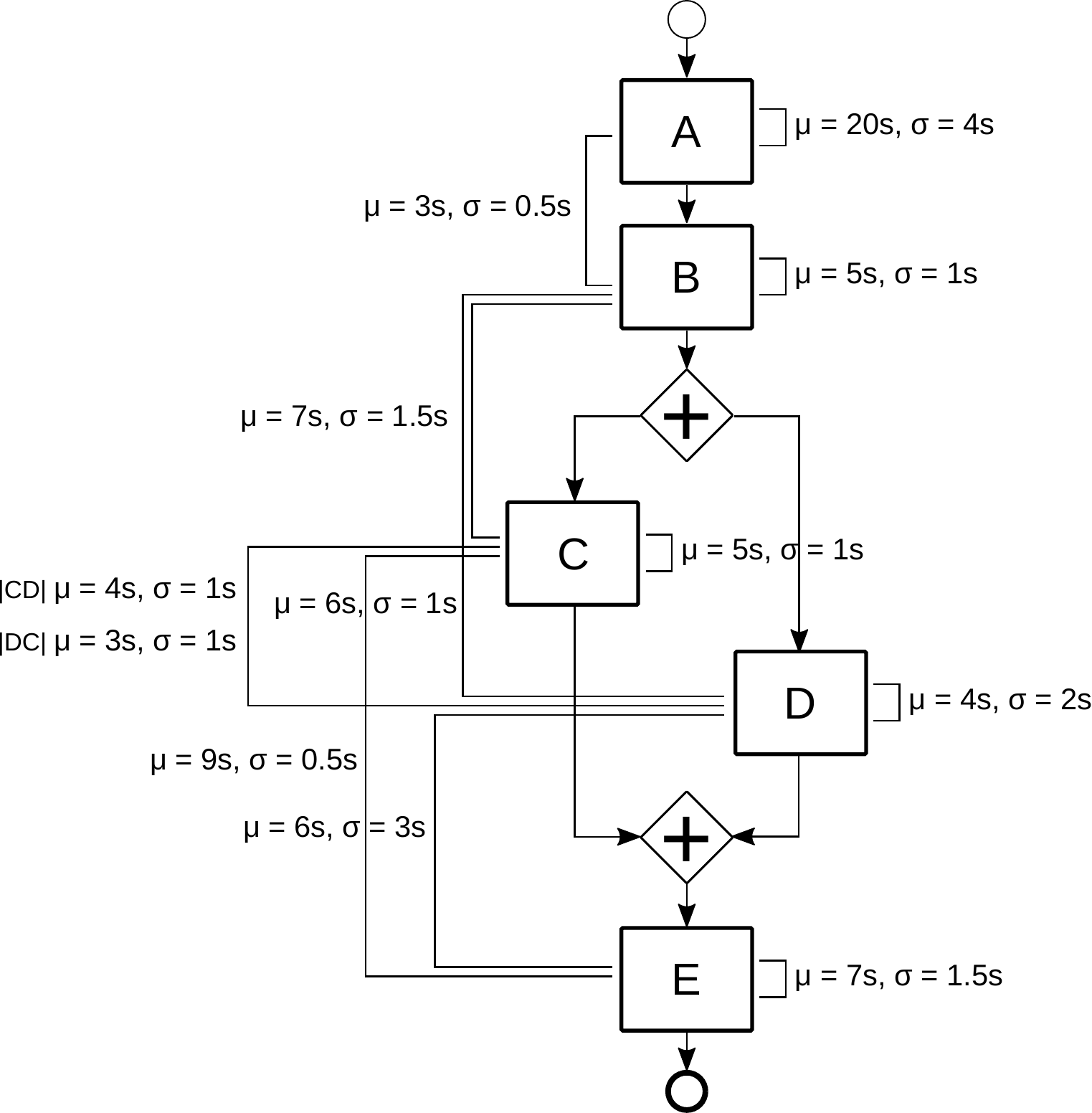}
\caption{Process Model Infused with Temporal Profile -- Example}\label{fig:running_ex}
\end{figure}

As can be seen in Fig. \ref{fig:running_ex}, the process model has been infused with a temporal profile capturing 
temporal distances and task durations on average and with the standard deviation.

Assume the following traces $t_1$ and $t_2$ to appear in an event stream of interest:\\
$t_1$ =
(($A_{start}$,0) ,($A_{end}$,19) ,($B_{start}$, 29)) \\\noindent  $t_2$ =
(($A_{start}$,0) ,($A_{end}$,20) ,($B_{start}$, 23),($C_{start}$, 24),($C_{end}$, 28),($B_{end}$, 29)).

Note that in order to illustrate 
the run through of Alg. \ref{alg:a1} and Alg. \ref{alg:a2}, the events are listed with their time spent
related to the start of the process instance in seconds. 

We set $\omega_{event}$ for $A$ and $B$ to 1 and for $|AB|$ to 2 to represent a
more sever deviation, $\phi$ to 1 for both instances and $\kappa_{event}$ to 2
for the task duration of $B$ and to 3 for all other distances. The events in
$t_1$ are in the correct order, so there are no costs for aligning this trace
to the process model.  The z-score for event $A$ is calculated when $A_{end}$
is detected, which yields 0.25. Since 0.25 is smaller than 3 ($\kappa$), no
temporal deviation cost is added. When $B_{start}$ is detected, the temporal
deviation cost is calculated based on the temporal distance of 10. This yields
a z-score of 14, which is greater than $\kappa$, increasing the cost for this
trace to 28, since $\omega$  is set to 2 for this distance. Assume that 7
seconds passed after $B_{start}$ has been performed and the algorithm has been
executed. Since that is greater than the average execution time of 6 seconds of
event $B$, the temporal deviation cost is calculated, which yields a z-score of
2. Because $\kappa$ is set to 2 for the task duration of $B$, the cost value is
increased by 6.  The current cost of this trace is therefore 34.

$t_2$ shows an deviating structure, since the execution of
$C$ has been executed before event $B$ finished, which results in move in the
alignment and by using the default conformance checking cost function yields
costs of 1.  The execution time duration of all events is fitting the process
model, as well as the temporal distances. There are no recordings for
the temporal distances $|AC|$ and $|CB|$, thus the temporal deviation costs
cannot be calculated for these distances.


\section{Evaluation}
\label{Sec:eval}

Algorithms \ref{alg:a1} and \ref{alg:a2}, together with the cost function for temporal deviations (cf. Def. \ref{def:cost}) are evaluated based on two real-world data sets. The
one is the BPI Challenge 2012 (BPIC2012 for short) \cite{bpic12}, which features a
process from the financial domain. This log has been chosen, since it provides
lifecycle attributes for at least a few number of events to calculate the temporal
distances. The feasibility of the approach is evaluated using this log.

Since no additional knowledge of the log is present to evaluate the results of
temporal deviation conformance checking, a second real-world is presented in
the evaluation. The second log \cite{figshare_bpm} features a business process
in the manufacturing
domain\footnote{\url{http://gruppe.wst.univie.ac.at/data/manufactoring_log.zip}}
and experts are evaluating the results afterwards to assess the applicability
of the approach.

\subsection{Financial Example}

The data set from the BPI Challenge 2012 contains $262200$ events from $13087$
process instances. Since only the process execution log is given by this data
set, we mainly focus on calculating the additional temporal deviation costs for an
alignment without the costs of structural conformance checking and assign
$\omega_x$ to 1 for every event. The events contain data elements describing the
values for a financial transaction as well as timestamps. A mandatory
requirement for the elaborated approach of this paper, is the availability of
lifecycle attributes of an event.

Out of the $23$ different tasks in the business process, only $6$ tasks have a
lifecycle logged with a start and an end event. This can be explained,
because the process consists of 3 inter twined sub-processes, but only
sub-process \texttt{W} contains the start and end event of activities, while the
other two \texttt{A} and \texttt{O} do not.

We opted for analyzing the data set as a whole instead of breaking it up into 3
different processes, because these sub-processes are inter-twined, therefore do
some temporal distances appear from sub-processes without start events.
The task duration does not change if other events happen during the
execution. Since no event stream of the data set is available, the data set
is split into a process execution log consisting of the first 80\% of process instances and an event stream
constructed using the other 20\% of the process instances, a similar approach to machine learning
algorithms \cite{alpaydin2020introduction}.

For Algorithm \ref{alg:a1}, the first $10469$ process instances are used as
a set to gather the temporal profile.

\begin{table*}[]
\centering
\begin{tabular}{l|l|l|l|l|l}
Name                            & Profile Size & $\mu$   & $\sigma$ & Min  & Max       \\ \hline
W\_Completeren aanvraag&18562&640.03&5883.48&0.77&244731.43 \\
W\_Nabellen offertes&18975&560.58&7302.64&0.95&243191.22 \\
W\_Valideren aanvraag&6493&1268.71&6098.85&1.1&238256.25 \\
W\_Afhandelen leads&4864&1012.62&9905.82&0.67&243739.82 \\
W\_Nabellen incomplete dossiers&9574&771.07&8052.62&1.03&239878.67 \\
W\_Beoordelen fraude&211&73.77&640.81&0.77&9240.84
\end{tabular}
\vspace{2mm}
\caption{BPIC 2012: Task Duration in seconds for all 6 events of 3271 process instances}
\label{tab:exec_bpic}
\end{table*}

As can be seen in Tab. \ref {tab:exec_bpic}, the task duration varies
to a great extent. Often the duration of an event takes seconds, other times the duration
spans several days. This can be attributed to the fact, that these events are
inter-twined with events from the other sub-processes. Hence the task
duration time of these events is depending on the time of the other events.
The duration cannot be calculated for the other sub processes, since
only end events are found in the process execution log.

As sub-process \texttt{W} contain start events, temporal distances can be
determined. We set $\kappa$ to 20 to filter out distances, that only have been
detected like 2\% of the time.

The remaining temporal distances can be seen in Tab.
\ref{tab:inter_bpic}.  Again, a wide range of time distances between the
events is detected. This is reasonable, because the task durations vary to a great extent as well.

\begin{table*}[]
\begin{tabular}{l|l|l|l|l|l}
Name                                                           & Profile Size & $\mu$     & $\sigma$  & Min   & Max        \\ \hline
A\_PREACCEPTEDW\_Completeren aanvraag&3819&22875.3&33214.99&3.55&156982.87 \\
W\_Completeren aanvraagW\_Nabellen offertes&4019&270672.33&308357.9&6.61&1206235.29 \\
W\_Nabellen offertesW\_Nabellen offertes&14955&258339.44&279374.88&1.4&2572740.11 \\
W\_Nabellen offertesW\_Valideren aanvraag&2640&233294.08&159396.81&7.2&617152.84 \\
W\_Completeren aanvraagW\_Completeren aanvraag&12686&79836.59&228186.05&1.47&2583000.9 \\
W\_Valideren aanvraagW\_Valideren aanvraag&2402&39583.82&112968.6&1.95&1289690.02 \\
A\_PARTLYSUBMITTEDW\_Afhandelen leads&3905&18670.85&30043.74&11.57&151792.25 \\
W\_Afhandelen leadsW\_Completeren aanvraag&2064&12132.28&20071.87&6.76&159723.98 \\
W\_Valideren aanvraagW\_Nabellen incomplete dossiers&1764&5108.96&8359.51&4.62&147659.13 \\
W\_Nabellen incomplete dossiersW\_Nabellen incomplete dossiers&7809&61731.99&112940.21&1.54&1117225.91 \\
W\_Nabellen incomplete dossiersW\_Valideren aanvraag&1421&34776.54&75051.87&9.97&614586.68 \\
W\_Afhandelen leadsW\_Afhandelen leads&940&3669.69&12485.41&1.46&160780.79
\end{tabular}
\vspace{2mm}
\caption{BPIC 2012: Temporal Distance in Seconds of the first 10469 process instances with $\kappa$ set to $200$}
\label{tab:inter_bpic}
\end{table*}

Algorithm \ref{alg:a2} is applied to the test set consisting of the remaining traces in the log. Even though
the algorithm has been designed for online execution, it is still possible
to calculate the temporal deviations in an offline setting. To achieve this all
events have been collected and have been sorted according to their timestamp.
Each of these events has then been injected into an event stream with random
intervals, but still using their original timestamp for the calculation. $\kappa_{event}$
has been set to 3 for all events, $\phi$ and all $\omega_x$ to $1$.

Out of $12650$ task duration, $12607$ events yielded a z-score below
$\kappa$, but $43$ yield a higher score. The maximum z-score of $3620.9$ has been found in
process instance \texttt{207263}.  As can be seen in Tab.
\ref{tab:exec_bpic}, \texttt{W\_Beoordelen fraude} is the event with a small
standard deviation. In this process instance the task is started at 09:00 on a
Thursday and finished the next day at 06:15. Thus a holiday does not seem
plausible and it is likely that something happened during this task. The
advantage of using this algorithm online would be the possibility to examine
the process instance immediately when enough time has past after the starting
event of this task. No process instance showed more than 2 deviations.

Out of $12654$ possible temporal distances, $12395$ do not deviate and $259$
do. The maximum deviations of one process instance is $5$, detected in $2$
process instances. Without domain experts, it is difficult to classify the
severeness of these deviations.

\subsection{Manufacturing Example}

The first example showed that the algorithm is returning good results with no
knowledge of the underlying process. For the manufacturing data
set\footnote{\url{http://gruppe.wst.univie.ac.at/data/manufactoring_log.zip}},
a process model with different $\omega_x$ and $\kappa_x$ is provided. An expert has been
assessing the results.

\begin{figure}[htb!]
\centering
\includegraphics[width=1.0\columnwidth]{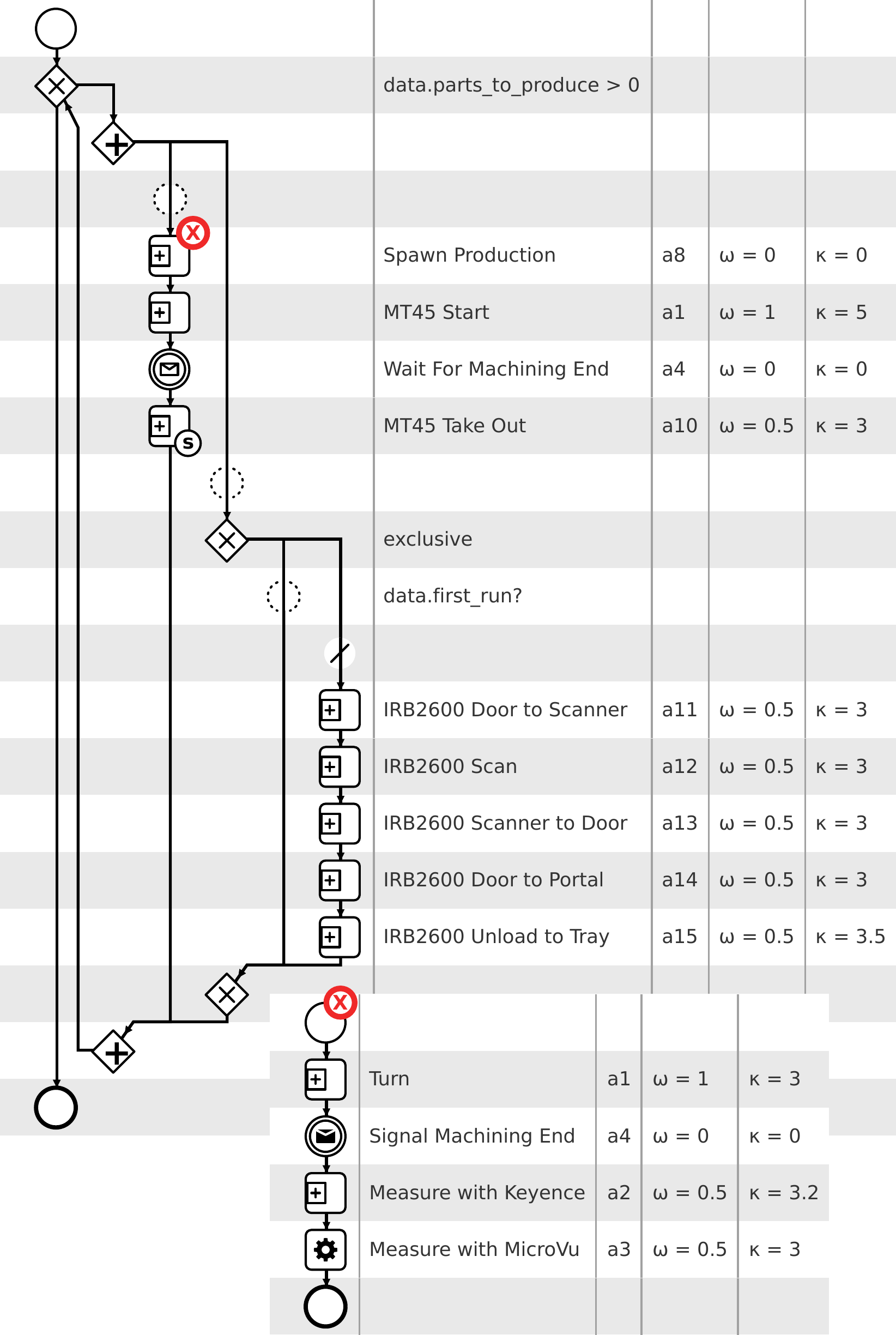}
\caption{Manufacturing of Parts}\label{fig:manu}
\end{figure}

Figure \ref{fig:manu} shows the manufacturing of a part\footnote{Note that the
tasks IDs a1 to a15, and b1 to b4, are neither continuous nor in sequence as
this version of the process is the result of several redesigns}. The main
process (tasks a1..15)ashows the interaction between different machines (EMCO
MT45 Turning Machine, ABB IRB2600 Robot, Keyence Optical Measurement Machine)
during production, while the sub-process \textcolor{red}{\circled{$\times$}} is
spawned for every single produced part, and tracks its full production
lifecycle. Every task has an $\omega$ assigned, values of $1$ signify normal
(high) importance. Values of $0$ signify that deviations can be ignored. In
addition to that, a $\kappa$ is assigned to each task, representing how many
standard deviations an observation can be distant to the mean of a distance.
Often 3 is used as a default value to detect outliers
\cite{crocker1986introduction}. A higher value allows for greater distant
distances, while a smaller value implies, that the distance has to be closer to
the mean.

The \textcolor{red}{\circled{$\times$}} sub-process is forked, i.e., the main
process does not wait for it to finish, but executes in parallel. The duration
between the b1 start event and the b1 end event should always be identical to
the duration between a8 start event and the a4 end event. For
$a1: \omega = 0$, as $a1$ forks a sub-process (without waiting for it to end). Hence
its duration is negligible.

Every occurrence of $\omega = 1$ is attributed to actual machining of
the part, which is expected to show small deviations. When the MT45 turning
machine closes its door and starts machining, only two cases can lead to a
deviation: A power failure or if raw material or the machining tool breaks.
Both of these cases are unlikely. Especially the latter is important.
If a tool or the raw material breaks, two things can happen: (1) the machine
triggers an emergency stop if any damage due to flying metal parts is
detected, or (2) the machining time is reduced, as there is no longer contact
between machining tool and raw material. Thus no friction occurs\footnote{This
can also be confirmed through lower power usage.}.

Thus, a first perceived application of temporal conformance checking is to trigger an emergency stop of the machines.

All tasks with the prefix IRB2600 are robot tasks (a10, a11, a12, a13, a14,
a15). The robot takes the finished parts out of the MT45 Turning Machine. This
can partly happen in parallel to the machining. As soon as the robot leaves
the confines of the MT45, the machine can start producing the next part. While
the MT45 is producing the next part, the robot (a12, b2) scans the part with
the help of a Optical Measurement Machine, and (a15) puts it on a tray which, in
turn, is placed on pallet with 50 other trays. All of the IRB2600 tasks are
expected to show small deviations, except for a15. As all the trays are on a
different position on the pallet, each operation should have a slightly
different duration, thus yielding the highest standard deviation of all tasks.
All robot tasks are also prone to extreme outliers, as the robot is subject to
an industrial safety mechanisms: whenever someone accidentally walks close to
the robot, a full emergency stop is triggered to avoid injuries. This can
indeed be observed in the data multiple times. Afterwards it is not always
trivial to restart the robot as sometimes it has to manually be moved into a
defined state, before the process can be continued.

As a second application, temporal conformance checking can be used to
automatically determine the significance of cases, in order to notify
personnel for solving the situation.

Table \ref{tab:exec_cdp} shows selected durations of different tasks. Temporal conformance checking
proves efficient in pointing out small deviations in the manufacturing
process, instantaneously at runtime. For the task duration, $27$
deviations of $1373$ distances and for the temporal distance, $47$
deviations of $1334$ distances have been detected  At one point the raw material
was running out. Thus the resulting part was missing some of its mass. Due to the
reduced weight, slightly different timings can be accurately pointed out in the
log. While this is probably also possible through other means (e.g., analyzing
the deviations in data from the measurement), the runtime temporal deviations
provide much faster and more universally applicable feedback.

\begin{table*}[]
\centering
\begin{tabular}{c|c|c|c|c|c}
Name                   & Profile Size & $\mu$ & $\sigma$ & Min   & Max   \\ \hline
IRB2600 Grip&30&40.66&6.3&37.32&74.03 \\
IRB2600 Extract&30&23.17&1.43&21.21&28.85 \\
IRB2600 Portal to Door&30&12.22&6.28&9.48&45.91 \\
IRB2600 Door to GS&30&12.47&0.56&11.48&13.85 \\
IRB2600 GS to Take&30&14.78&0.47&13.96&15.81 \\
IRB2600 Take to GS&30&10.5&1.1&9.69&15.73 \\
IRB2600 GS to Door&30&11.49&0.7&10.1&13.12 \\
IRB2600 Door to Scanner&30&13.75&1.62&11.19&18.2 \\
IRB2600 Scan&30&29.59&1.82&26.08&32.59 \\
IRB2600 Scanner to Door&30&12.89&0.98&11.19&14.84 \\
IRB2600 Door to Portal&30&10.96&1.47&9.56&18.03 \\
IRB2600 Unload to Tray&30&20.86&9.6&17.4&69.16
\end{tabular}
\vspace{2mm}
\caption{Manufacturing: Task Duration in Seconds of first 30 process instances in the data set.}
\label{tab:exec_cdp}
\end{table*}

\textbf{Threats to validity}: Even though real world data sets are used in this
evaluation there are still potential problems. One concern is the volume and
velocity of data in the event stream. While both of these data sets are rather
small and can easily fit into the memory of a modern computer, there could be
performance issues when dealing with a real infinite event stream.  Another
important aspect is the need for an expert to receive satisfying results. Even
though deviations have been detected in the finance example, it is hard to
argue automatically if these deviations are a real concern for the process
instance or are negligible. It should also be noted, that a process execution
engine put events into the event stream, therefore the order of the event stream reflected
the order of the execution. If the event stream is not ordered, a preprocessing step has to be
inserted.

%
%
\section{Related Work}
\label{Sec:rel}

There is a plethora of established offline techniques for process discovery
and process conformance checking \cite{DBLP:books/sp/Aalst16,rozinat2008conformance,van2012replaying}. Offline
techniques calculate the results after the process instances have been
finished. Offline techniques tend to yield more precise results, since the complete
data set is available. However, a repair of a broken instance at runtime is not possible anymore.

In \cite{eder2001managing} identified the impact of a temporal framework,
introduced and described the modeling of  time constraints in workflow models.

An online approach for conformance checking can be found in
\cite{van2017online, burattin2014control}. Online approaches are being executed while the process is
running and use an event stream as data set.  In \cite{van2017online} prefix alignments are
introduced, which can be calculated at run-time incrementally. The method also
introduces an approximation which allows a better memory efficiency. 

Even in an offline environment, finding the best alignment with the least
costs constitutes a heavy computational task. \cite{sani2020conformance} argues that for
big process execution logs, traditional conformance checking cannot be used,
since the computation is taking too long. The paper suggests an approximation
to reduce the computation time drastically, still yielding a satisfying result.
To accomplish that, only a subset of potential event sequences for a process model
are considered.

Tesseract \cite{richter2017tesseract}, focuses on temporal deviations to detect
concept drifts in a business process. It is capable of detecting sudden drifts,
i.e., the process model is changed instantly, and incremental drifts, i.e., the
process model is changed in a small way like an additional event or different
execution time.  This method aims at detecting concept drifts and does not
quantify temporal deviations of a single instance when compared to a process model.

\cite{9018187}  does not use a process execution log,
but a data stream of all the data elements of a business process. These data
elements are stored in time sequences and the behavior of these time sequences
is evaluated. Instead of comparing the complete time sequences with each
other as in \cite{stertzbpm} using Dynamic Time Warping \cite{berndt1994using},
smaller chunks are compared to check whether or not the behavior changes between two events,
i.e., data values increase instead of decrease.

A Temporal Network Representation (TNR) is introduced in
\cite{senderovich2017temporal}. This method detects the temporal relation
between events and establishes a method to discover unbiased models using
TNR-based inductive mining.

The likelihood of an event occurrence is usually not taken into account for conformance checking.
Stochastic conformance checking is established in \cite{leemans2019earth}. It enriches petri nets
with stochastic values and introduces a stochastic language, to describe the likelihood of a specific transition firing.
The Earth Mover"s Distance is used for calculating the conformance. 

Multi-perspective conformance checking
\cite{DBLP:journals/computing/MannhardtLRA16} introduces multi view conformance
checking based on data elements, like resource and time. In their approach
conformance is checked based on constraints for the data elements contrary to a
previously mined temporal profile.

Temporal anomaly detection \cite{rogge2014temporal}, aims at finding deviations
in the execution time duration using a Bayesian model as well as distinguishing
between temporal deviations and measurement errors. Contrary to our approach,
outliers are only pointed out, but not quantified in term of fitness costs.


%
\section{Conclusion}
\label{Sec:concl}

This paper introduces temporal conformance checking in order to detect and quantify temporal
deviations for task durations and temporal distances between events. The approach can be executed in an offline and online
setting. For an offline setting, a process execution log can be split in training and test sets. Based on the training set and the process model, a temporal profile is calculated. In addition, the tasks in the process model can be annotated with their significance for temporal conformance. The temporal profile and the model can then be compared to the test set (offline) or an even stream of interest (online). 

The evaluation is conducted with two real-world data sets. The data set from the financial domain demonstrated the feasibility of the approach for an offline setting and without expert knowledge. The data set from the manufacturing domain showed the applicability of the approach in an online setting. Moreover, as the results could be checked with an expert, two applications for the proposed temporal conformance checking were identified, i.e., triggering emergency machine stops and based on the significance of the deviations notifying personnel. 

A drawback of this approach is the requirement of lifecycles attached to
events, in order to distinguish start and end events. Without this, the distance within
events cannot be detected at all and the distance between events can only be
guessed. 

For future work, we plan an online visualization which aims at making
temporal deviations easily detectable by experts. A more in-depth analysis on
time distances for parallel tasks is planned as well as dynamically calculating
weights assigned to a task, to better reflect the impact of certain deviations.

\section*{Acknowledgment} This work has been partly funded by the
Austrian Research Promotion Agency (FFG) via the ``Austrian Competence Center
for Digital Production'' (CDP) under the contract number 854187. This work has
been supported by the Pilot Factory Industry 4.0, Seestadtstrasse 27, Vienna,
Austria.

\bibliographystyle{splncs03}
\bibliography{bib}
\end{document}